\documentclass[11pt,preprint]{aastex}
\usepackage{geometry}                % See geometry.pdf to learn the layout options. There are lots.
\usepackage{graphicx}
\usepackage{amssymb}
\usepackage{epstopdf}
\begin{document}
\title{Titan's transport-driven methane cycle}
\author{Jonathan L. Mitchell}
%\date{}                                           % Activate to display a given date or no date

%\maketitle

\begin{abstract}
The strength of Titan's methane cycle, as measured by precipitation and evaporation, is key to interpreting fluvial erosion and other indicators of the surface-atmosphere exchange of liquids.  But the mechanisms behind the occurrence of large cloud outbursts \citep{Schaller_etal06} and precipitation \citep{Turtle_etal09,Turtle_etal11} on Titan have been disputed.  A gobal- and annual-mean estimate of surface fluxes indicated only 1\% of the insolation, or $\sim$0.04 W/m$^2$, is exchanged as sensible and/or latent fluxes \citep{mckay1991greenhouse}.  Since these fluxes are responsible for driving atmospheric convection, it has been argued that moist convection should be quite rare and precipitation even rarer, even if evaporation globally dominates the surface-atmosphere energy exchange \citep{griffith2008titan}.  In contrast, climate simulations that allow atmospheric motion indicate a robust methane cycle with substantial cloud formation and/or precipitation \citep{rannou2006latitudinal,mitchell2006dynamics, mitchell2008drying, mitchell2009impact, mitchell2011locally, schneider2012polar}.  We argue the top-of-atmosphere radiative imbalance -- a readily observable quantity -- is diagnostic of horizontal heat transport by Titan's atmosphere, and thus constrains the strength of the methane cycle.  Simple calculations show the top-of-atmosphere radiative imbalance is $\sim$0.5-1 W/m$^2$ in Titan's equatorial region, which implies 2-3 MW of latitudinal heat transport by the atmosphere.  Our simulation of Titan's climate suggests this transport may occur primarily as latent heat, with net evaporation at the equator and net accumulation at higher latitudes.  Thus the methane cycle could be 10-20 times previous estimates.  Opposing seasonal transport at solstices, compensation by sensible heat transport, and focusing of precipitation by large-scale dynamics could further enhance the local, instantaneous strength of Titan's methane cycle by a factor of several.  
\end{abstract}

Titan's transport-dominated methane cycle is fundamentally distinct from the hydrological cycle of Earth's tropical zone. Earth's hydrology is primarily driven by local radiative imbalance at the surface that amounts to $\sim$115 W/m$^2$ of evaporation occurring over the tropics.  An additional $\sim$40 W/m$^2$ of evaporation results from horizontal heat transport away from the equator by the motion of the atmosphere-ocean system \citep{trenberth2003covariability}.  
%In contrast, Titan's methane cycle is driven almost entirely by horizontal heat transport by the atmosphere.  This process is not accounted for in previous estimates of Titan's hydrology based on one-dimensional radiative-convective modeling (McKay).  
Following conventions from the literature on Earth's climate energetics \citep{trenberth2003covariability}, we identify the role of energy transport by the atmosphere in Titan's global energetics.  Vertically integrated quantities are defined $\tilde{M} = \int_0^\infty \rho M dz = \frac{1}{g} \int_0^{p_s} M dp$.  The total energy of the atmosphere, after vertical integration, is 
\begin{eqnarray}
A_E &=& \tilde{k} + \widetilde{c_p T} + \tilde{\Phi}_s + \widetilde{Lq} \\
	&\approx& D_E + L_E = M_E \ ,
\end{eqnarray}
with kinetic energy $K_E = \tilde{k}$ (assumed to be negligible), internal energy $\widetilde{c_p T}$, potential energy $\tilde{\Phi}_s$ (which combine to give dry static energy $D_E = \widetilde{c_p T} + \tilde{\Phi}_s$) and latent energy $L_E=\widetilde{Lq}$.  The atmospheric energy is very nearly equivalent to the vertical integral of the moist static energy, $M_E = D_E+L_E = \tilde{h}$, with $h=c_pT +gz + Lq$. 

The thermodynamic equation,
\begin{eqnarray}
\frac{\partial D_E}{\partial t} &=& -{\bf \nabla\cdot F_{D_E}} + Q_1 \ ,
\end{eqnarray}
and the moisture equation,
\begin{equation}
\frac{\partial L_E}{\partial t} = -{\bf \nabla\cdot F_{L_E}} - Q_2 \ ,
\end{equation}
combine to give the moist static energy equation,
\begin{equation}
\frac{\partial M_E}{\partial t} = -{\bf \nabla\cdot F_{M_E}} + Q_1 - Q_2 \ ,
\end{equation}
where
\begin{eqnarray}
{\bf F_{D_E}} &=& \widetilde{{\bf v} c_p T} + \widetilde{{\bf v}\Phi} \\
{\bf F_{L_E}} &=& \widetilde{{\bf v} Lq} \\
{\bf F_{M_E}} &=& {\bf F_{De}} + {\bf F_{Le}} \\
Q_2 &=& L(\tilde{P} - \tilde{E}) \\
Q_1 &=& R_T - R_s + H_s + L\tilde{P} \\
 	&=& R_T + F_s + Q_2 \ . 
\end{eqnarray}
$Q_1$ is the column-integrated diabatic heating in W/m$^2$.  Atmospheric heating can arise from radiative imbalances at the top-of-atmosphere, $R_T$, and surface, $R_S$, from surface sensible heat fluxes, $H_S$, or the release of latent heat by precipitation, $L\tilde{P}$.  The column-integrated latent heating, $Q_2$, is the difference of precipitation and evaporation, $L(\tilde{P}-\tilde{E})$.  

In a steady-state, there is a balance between column heating, $Q_1-Q_2$, and horizontal divergence of moist static energy, ${\bf \nabla\cdot F_{M_E}}$. In an atmosphere with no horizontal heat transport (as is true in single-column, radiative-convective models), this balance reduces to $Q_1=Q_2$, which further reduces to $F_s = R_T$ if the atmosphere does not store heat.  In this case, the net surface flux $F_s = LE_s+H_s-R_s$ is equal and opposite to the top-of-atmosphere radiative imbalance $R_T=S-OLR$, i.e., a heat source/sink at the top of the column must be offset locally by a sink/source at the bottom.\footnote{Our definition of $S$ is the net solar flux, and includes the reduction due to local albedo.} Global equilibrium requires the average of $R_T$ to be zero, therefore on average the surface radiative imbalance $R_s$ is offset locally by turbulent exchange of latent heat $LE_s$ and sensible heat $H_s$ with the atmosphere.  Based on these arguments, a one-dimensional radiative-convective model of Titan with $R_s \simeq 0.04 \mbox{ W/m}^2$ \citep{mckay1991greenhouse} is often invoked as a constraint on the evaporative surface flux, $LE_s \leq R_s$, a severe limitation on the strength of Titan's methane cycle \citep{griffith2008titan}.  We argue that horizontal heat transport by the atmosphere may dramatically increase this limit.  

\begin{figure*}[htbp]
\begin{center}
\plotone{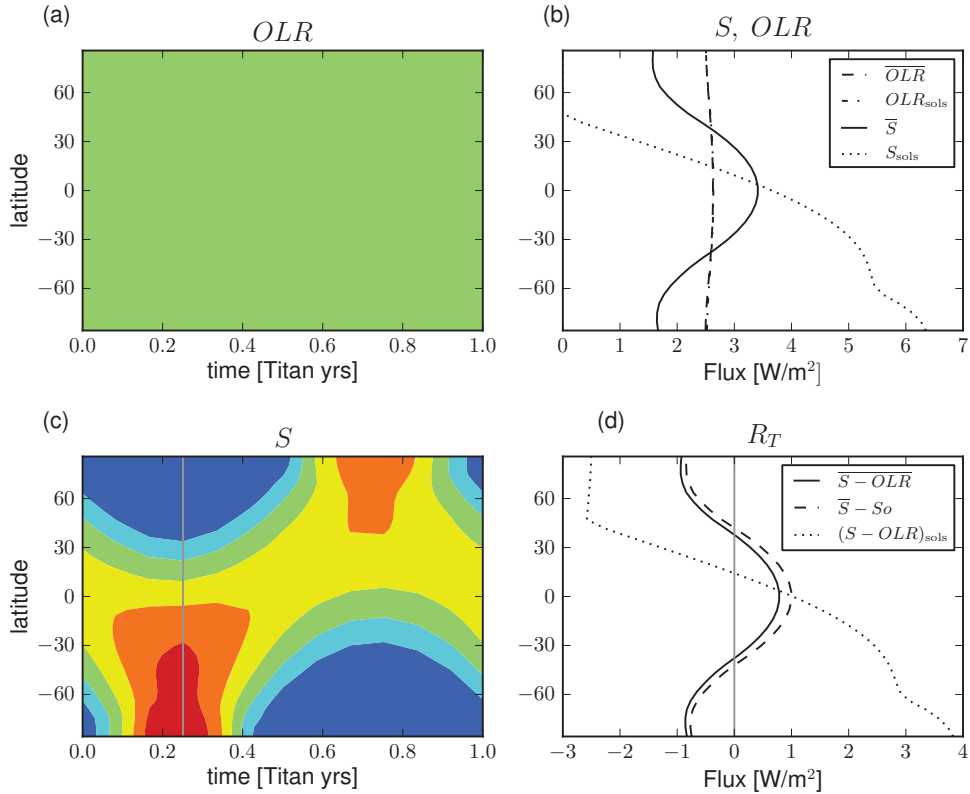}
\caption{Titan's top-of-atmosphere (TOA) radiative fluxes are show separately as outgoing longwave radiation ($OLR$) and insolation ($S$) as a function of seasons in (a \& c) on the same color scale, and as annual-means or instantaneously as indicated in (b \& c).  Titan's $OLR$ (a) is nearly independent of latitude and time, while the insolation (c) varies substantially with season.  In the annual mean, insolation is strongly peaked at and symmetric about the equator (solid line).  Since the $OLR$ is flat, there is a net imbalance in TOA radiation (c), with $\sim$0.3 W/m$^2$ of heating at the equator and $\sim$0.5 W/m$^2$ cooling at high latitudes.  The insolation at southern summer solstice (dotted line) is peaked at the summer pole and gives rise to a net TOA imbalance of up to 2 W/m$^2$.}
\label{TOA}
\end{center}
\end{figure*}

\begin{figure*}
\plotone{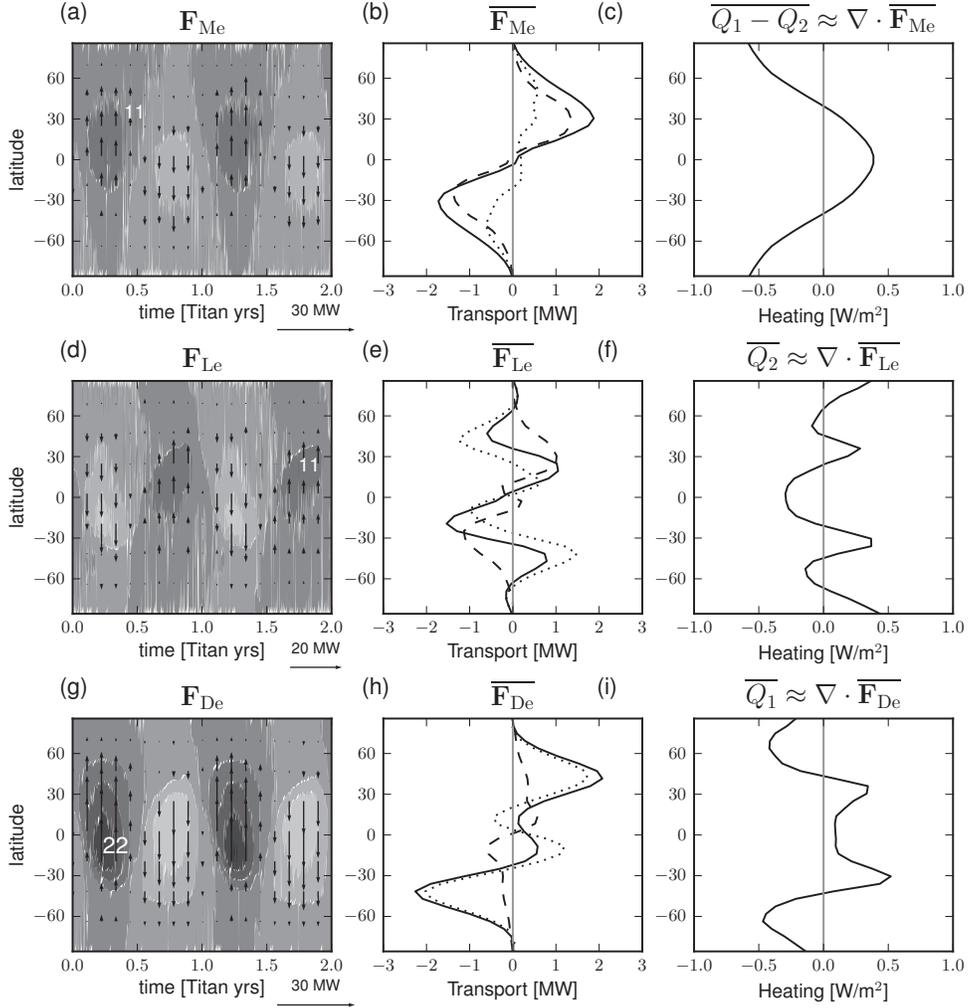}
\caption{Our model of Titan's climate dynamics as a function of latitude and season (a) indicates peak transport of 11 MW.  In the annual-mean (b) the transport needed is $\sim$2 MW to offset the TOA radiative imbalance (c \& Figure \ref{TOA}).  Time-dependent (dashed line in b) and mean (dotted line) transport both contribute to the moist static energy transport.  The latent heat transport (d) peaks after solstices at 11 MW, and is offset by dry-static energy transport (g) in the opposing direction.  The total moist static energy transport in the equatorial region (b) is dominated by latent heat transport (e) while dry-static transport is nearly negligible.  Divergent latent energy transport in the equatorial region (e) is balanced by net evaporation near the equator and accumulation in mid-latitudes (f).}
\label{moist_fluxes}
\end{figure*}

In order to derive a constraint on the strength of Titan's methane cycle in the presence of meridional heat transport by the atmosphere, first we observe Titan's $OLR$ is essentially ``flat'' (independent of latitude) and ``steady'' (independent of time) \citep[Figure \ref{TOA}a]{li2011energy}.  Time-independence results from the long radiative cooling time of Titan's middle troposphere, where the atmosphere radiates to space.  However, long thermal times do not account for a flat $OLR$.  Indeed without atmospheric heat transport, Titan's $OLR$ would exactly balance the annual-mean insolation which is strongly peaked at the equator and decreases poleward (Figure \ref{TOA}b, solid line).  Instead, Titan's flat $OLR$ (Figures \ref{TOA}a,b) is direct evidence for poleward energy transport by the atmosphere (Figure \ref{moist_fluxes}).  Since the insolation $S$ is a function of both space and time and we require global energy balance on long timescales (i.e., constant $A_E$), the top-of-atmosphere radiative imbalance $R_T \approx S-S_o$ with $S_o$ the global- and time-mean insolation (Figure \ref{TOA}d, dashed line).  In this case, $R_T$ is offset either by surface energy storage or by horizontal transport of moist static energy.  The former option is ruled out by the solid, low-heat-capacity surface of Titan which prevents energy from being transported or stored for long periods.  Thus, we infer a divergent flux of moist static energy from regions of positive $R_T$ to regions of negative $R_T$ (Figures \ref{TOA}d and \ref{moist_fluxes}b).  

The implied heat flux from $R_T$ is in general a combination of latent and dry static energy (Figure \ref{moist_fluxes}), however we argue the imbalance itself provides a rough estimate of the magnitude of column cooling by latent heat transport out of the tropics (the ``drying rate'' do to evaporation),
\begin{equation}
%F_{\rm lat} & \sim & 1 \mbox{ m/yr} \ /\ 3.15\times10^7 \mbox{ s/yr} \ * \ 10^3 \mbox{ kg/m}^2 \ * \ 2.5\times10^6 \mbox{ J/kg} \\
%F_{\rm E} & \sim & 100 \left(\frac{P}{1.25 \mbox{ m/yr}}\right)\mbox{ W/m}^2 \\
Q_2 \sim L\tilde{E} \sim0.5 \left(\frac{E}{7 \mbox{ cm/yr}}\right)\mbox{ W/m}^2 \ ,
\end{equation}
which is in fact the value obtained in our simulation (Figure \ref{moist_fluxes}f).  

Our arguments thus far are based on the assumptions of top-of-atmosphere radiative imbalance, surface energy balance, and the dominance of poleward latent heat transport.  Further insight into Titan's methane cycle can be gained through global circulation model (GCM) simulations.  In the annual mean, the atmospheric heat transport implied by the TOA radiative imbalance is from the equator to higher latitudes.  However, simulations \citep{rannou2006latitudinal,mitchell2006dynamics, mitchell2008drying, mitchell2009impact, mitchell2011locally, schneider2012polar} and cloud observations \citep{brown2010clouds, rodriguez2011titan, turtle2011seasonal} indicate the transport seasonally reverses from a globally northward phase to a southward phase, with large cancellation between the two phases in the annual mean.  A compensation between dry-static and latent transports may further enhance the strength of Titan's methane cycle, as we now demonstrate.

In the remainder of this paper, we substantiate the above arguments and draw attention to differences between Titan's methane cycle and Earth's hydrological cycle by comparing GCM simulations of Titan with NCEP reanalysis \citep{kalnay1996ncep} for Earth.  We focus our attention on a ``moist'' version of Titan, i.e., one with an unlimited and pure supply of surface methane, in order to establish the potential strength of the methane cycle.  We then discuss the effects a limited surface supply and the presence of liquid hydrocarbon mixtures would have on our results.

Figure \ref{moist_fluxes} shows a breakdown of the energetics of our GCM simulation of Titan reported in \cite{mitchell2011locally}.  The flux of moist static energy seasonally reverses with peak magnitudes of $\sim$11 W/m$^2$ (\ref{moist_fluxes}a).  In the annual mean, there is flux divergence at the equator and convergence in mid-latitudes, with peak fluxes of $\sim$2 MW (\ref{moist_fluxes}b).  The annual-mean net column heating $\overline{Q_1-Q_2}$ (top-right panel) is positive at the equator and negative at high latitudes, indicating an excess of $\sim$0.5 W/m$^2$ insolation at the equator and an excess of $>0.5$ W/m$^2$ OLR near the poles.  The divergence of moist static energy offsets the net column heating.

At this point, we encounter the first feature of Titan's climate that distinguishes it from Earth's tropical climate.  When we break the components of the moist static energy transport into latent (Figure \ref{moist_fluxes}d,e,f) and dry static (Figure \ref{moist_fluxes}g,h,i) transport, we find that the annual-mean heat flux divergence at low latitudes is dominated by latent energy fluxes.  As a result, the equatorial region experiences net drying (negative $\overline{Q_2}$; Figure \ref{moist_fluxes}f) of $\sim$0.5 W/m$^2$, while mid-latitudes have net accumulation of a similar magnitude, an effect already well documented in several Titan climate models \citep{rannou2006latitudinal,mitchell2006dynamics, mitchell2008drying, mitchell2009impact, mitchell2011locally, schneider2012polar}.  This confirms our earlier assumption, that the magnitude of the equatorial drying rate is set by the horizontal latent energy transport required to keep Titan's OLR ``flat''.  

A second distinguishing feature of Titan's climate follows by dividing the annual-mean latent energy transport into components due to the time-mean and variations as $L_E = \overline{L_E}+L_E'$, such that for instance $\overline{{\bf F_{\rm Le}}} = L\widetilde{\overline{v}\overline{q}} + L\widetilde{\overline{v'q'}}$, we find that the majority of the energy transport occurs due to the time-dependent seasonal cycle ($L\widetilde{\overline{v'q'}}$; dashed line in Figure \ref{moist_fluxes}e) rather than by the annual-mean component ($L\widetilde{\overline{v}\overline{q}}$; dotted line).  In fact, the mean component is convergent at the equator, but the seasonal divergence dominates.  In Earth's tropics, by contrast, the atmosphere accomplishes moist static energy transport primarily by the mean component \citep{trenberth2003seamless}.  

Our analysis reveals a third important feature of Titan's atmospheric energetics: a large compensation between dry-static and latent energy transports during solsticial conditions (Figure \ref{moist_fluxes}a,d,g).  In fact, instantaneous latent energy fluxes oppose the moist static energy flux (compare Figures \ref{moist_fluxes}a and d), implying that moisture is being fluxed towards the summer pole.  Instantaneous dry-static transported from the summer to the winter hemisphere carries is roughly double the latent transport (Figure \ref{moist_fluxes}g), and in the annual mean is moderately convergent at the equator (Figure \ref{moist_fluxes}h, solid line).  This compensation is also a feature of Earth's annual-mean tropical circulation \citep{trenberth2003seamless}, but in contrast Earth's quasi-steady tropical circulation converges moisture to the equator and diverges dry-static energy towards the poles.  
%A second type of compensation between surface sensible heat flux and evaporation is possible on Titan (so-called ``negative Bowen ratio'' conditions; Figure \ref{surface}a).  Compensating surface fluxes are not a feature of Earth's tropical, large-scale surface energetics, but they do occur at high latitudes (Figure \ref{surface}b).  The level of flux compensation is intrinsically uncertain in models, being dependent on poorly constrained processes such as boundary layer turbulence in statically stable conditions.  
Based on our simulations, it seems plausible that Titan's seasonal cycle enhances the instantaneous methane transport and precipitation by at least a factor of five, or $\sim$10 W/m$^2$.

\begin{figure*}
\plotone{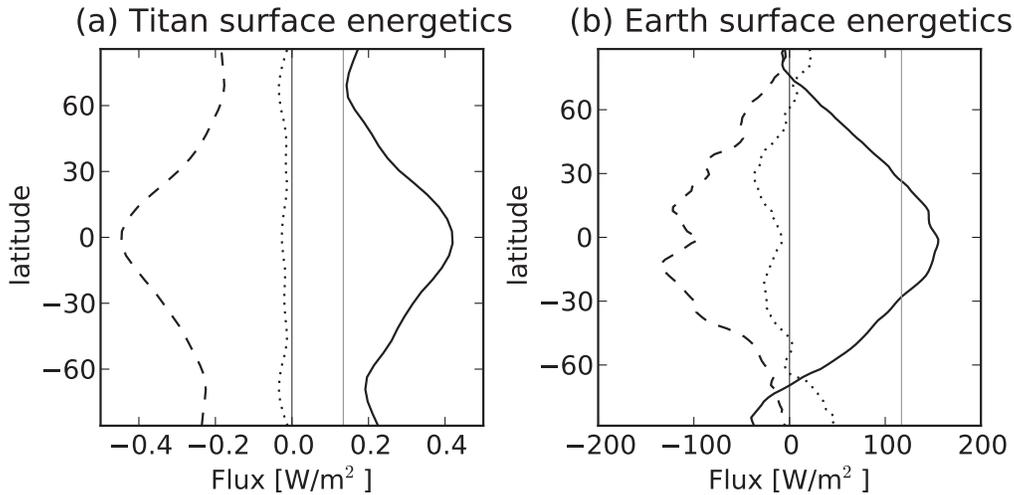}
\caption{Annual-mean net surface radiative fluxes, which are responsible for driving evaporation and convection, for our Titan model with latitudinal heat transport (a, solid line) exceed radiative fluxes in a model without heat transport and forced by the global-mean insolation (thin gray line) by a factor of six.  In addition, there is a compensation between dry static energy fluxes into the surface (dotted line) and evaporative cooling (dashed line) which further enhances the methane cycle.  In contrast, Earth's surface energetics (b) is dominated by direct solar heating, as is indicated by the net surface radiative imbalance in a model calculation without horizontal heat transport (thin gray line) as compared to the imbalance in NCEP reanalysis (solid line).  Latitudinal transport accounts for a $\sim$30\% enhancement in the equatorial radiative imbalance.  Both sensible (dotted line) and evaporative (dashed line) fluxes act to cool Earth's surface, except in very high latitudes.}
\label{surface}
\end{figure*}

\begin{figure*}
\plotone{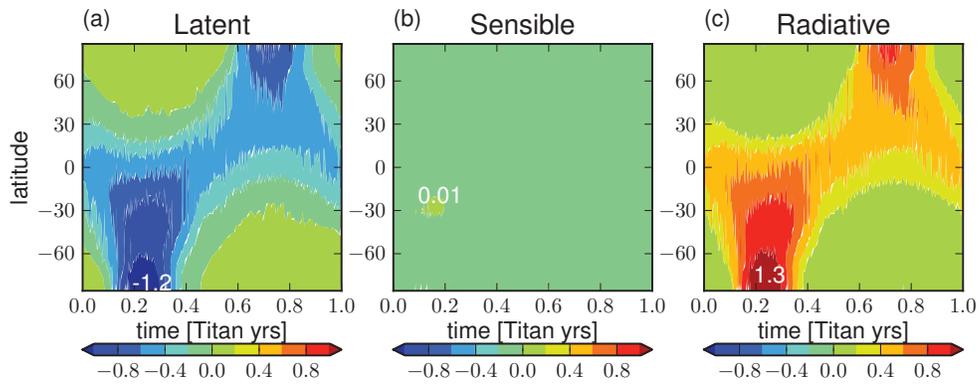}
\caption{The zonal-mean surface energetics of our model is dominated by a local balance between evaporative cooling (left) and radiative heating (right), with a relatively small role for sensible heat fluxes (center). Local, instantaneous surface radiative imbalance peaks at the summer pole, exactly when the top-of-atmosphere imbalance peaks (Figure \ref{TOA}).}
\label{sfc_flux_vs_time}
\end{figure*}

The surface radiative flux imbalance is the energy available to drive turbulent exchange of sensible (dry-static) and latent heat fluxes, and is often cited as an upper-limit to the evaporative flux in Titan's climate system \citep{griffith2008titan,williams2012energy}.  A comparison between annual-mean surface radiative imbalance (Figure \ref{surface}) from our ``dynamic model'' with horizontal heat transport (solid line) and in radiative-convective equilibrium with the annual- and global- mean insolation (thin gray line) demonstrates the influence of atmospheric circulation on Titan's methane cycle and Earth's hydrological cycle.\footnote{In the case of Earth, radiative-convective equilibrium was calculated with a single-column version of the CCM3 model forced by the annual- and global-mean insolation.  NCEP reanalysis is used in lieu of a ``dynamic model''.}  For Titan, the surface radiative imbalance in the dynamic model is everywhere larger than the radiative-convective model surface imbalance.  The modeled instantaneous fluxes can be a factor of several larger than this (Figure \ref{sfc_flux_vs_time}) primarily due to the strong seasonal cycle in the TOA radiative imbalance (Figure \ref{TOA}). 
By contrast, Earth's surface radiative imbalance is dominated by ``local'' heating that does not involve latitudinal heat transport (Figure \ref{surface}b, thin gray line).  Comparing with NCEP fluxes, we see that transport accounts for a $\sim$30\% enhancement of radiative imbalance at the equator (solid line).  This dominance of ``local'' energetics in Earth's climate may be the origin of misconceptions about Titan's climate that lead to large underestimates of the potential strength of the methane cycle.  To reiterate, Titan's ``flat'' OLR (Figure \ref{TOA}) is direct evidence for 2-3 MW of latitudinal heat transport by the atmosphere (Figure \ref{moist_fluxes}b), our numerical experiments suggest much of it could be transported as latent energy (Figure \ref{moist_fluxes}e), and therefore Titan's methane cycle has the potential to be much stronger than is estimated based on global average surface energetics.

%A more relevant constraint on Titan's methane cycle is the ability of the atmosphere to cool radiatively, since this is the only path to rid the atmosphere of latent heat released during precipitation.  The steady-state, global- and annual-mean precipitation $\overline{LP}$ is equivalent to the global- and annual-mean radiative imbalance, $.  

There exist a few mechanisms not accounted for in our analysis that could throttle Titan's methane cycle.  Firstly and most importantly, there is apparently a limited supply of liquid methane at the surface \citep{lorenz2008titan}.  Although precipitation does occur in the equatorial region \citep{Turtle_etal11}, dry conditions are expected to persist there in the present climate \citep{rannou2006latitudinal,mitchell2006dynamics, mitchell2008drying, mitchell2009impact, schneider2012polar}.  Sensitivity experiments with our model demonstrate that as the global methane supply is reduced, the latent heat transport in the equatorial region reduces in-kind and dry static transport increases to offset the reduction \citep{mitchell2008drying}.  Secondly, liquid ethane may be well-mixed in liquid methane at the surface, and if so it would suppress the saturation vapor pressure of methane at a level proportional to the unknown mixing ratio of the two hydrocarbons in surface reservoirs.  We tested this effect in our two-dimensional climate simulations by artificially suppressing the evaporation of surface methane \citep{mitchell2006dynamics, mitchell2009impact}.  Our results indicate that while there is a substantial reduction in the annual-mean evaporation rate relative to the ``moist'' case, the instantaneous latent heat transport at solstices is only reduced by a factor of 2-3, i.e., to 7 MW \citep{mitchell2008drying}.  Local evaporation and accumulation in the annual-mean in this scenario are of order 0.5-1 W/m$^2$, similar to the ``moist'' case (Figure \ref{moist_fluxes}f). 

In summary, Titan's nearly constant OLR, seasonal lower tropospheric circulation, and compensation between dry-static and latent energy transports allow a vigorous methane cycle despite weak, global-mean radiative forcing.  Titan's storms are triggered in large-scale updrafts \citep{mitchell2006dynamics, mitchell2008drying, mitchell2009impact, schneider2012polar} and waves \citep{mitchell2011locally}.   Large cloud outbursts \citep{Schaller_etal06} and precipitation events \citep{Turtle_etal09, Turtle_etal11} occur as the result of the global transport and focusing power of Titan's atmospheric circulation, with significant implications for the interpretation of surface features ascribed to the presence of flowing and evaporating liquids.

\clearpage

\clearpage
\begin{thebibliography}{20}
\expandafter\ifx\csname natexlab\endcsname\relax\def\natexlab#1{#1}\fi

\bibitem[{Brown {et~al.}(2010)Brown, Roberts, \& Schaller}]{brown2010clouds}
Brown, M., Roberts, J., \& Schaller, E. 2010, Icarus, 205, 571

\bibitem[{Griffith {et~al.}(2008)Griffith, McKay, \& Ferri}]{griffith2008titan}
Griffith, C., McKay, C., \& Ferri, F. 2008, The Astrophysical Journal Letters,
  687, L41

\bibitem[{Kalnay {et~al.}(1996)Kalnay, Kanamitsu, Kistler, Collins, Deaven,
  Gandin, Iredell, Sana, White, Woollen, {et~al.}}]{kalnay1996ncep}
Kalnay, E., Kanamitsu, M., Kistler, R., Collins, W., Deaven, D., Gandin, L.,
  Iredell, M., Sana, S., White, G., Woollen, J., {et~al.} 1996, Bull. Amer.
  Meteor. Soc.

\bibitem[{{Li} {et~al.}(2011){Li}, {Nixon}, {Achterberg}, {Smith}, {Gorius},
  {Jiang}, {Conrath}, {Gierasch}, {Simon-Miller}, {Michael Flasar}, {Baines},
  {Ingersoll}, {West}, {Vasavada}, \& {Ewald}}]{li2011energy}
{Li}, L., {Nixon}, C.~A., {Achterberg}, R.~K., {Smith}, M.~A., {Gorius},
  N.~J.~P., {Jiang}, X., {Conrath}, B.~J., {Gierasch}, P.~J., {Simon-Miller},
  A.~A., {Michael Flasar}, F., {Baines}, K.~H., {Ingersoll}, A.~P., {West},
  R.~A., {Vasavada}, A.~R., \& {Ewald}, S.~P. 2011, Geophysical Research
  Letters, 38, 23201

\bibitem[{Lorenz {et~al.}(2008)Lorenz, Mitchell, Kirk, Hayes, Aharonson,
  Zebker, Paillou, Radebaugh, Lunine, Janssen, {et~al.}}]{lorenz2008titan}
Lorenz, R., Mitchell, K., Kirk, R., Hayes, A., Aharonson, O., Zebker, H.,
  Paillou, P., Radebaugh, J., Lunine, J., Janssen, M., {et~al.} 2008,
  Geophysical Research Letters, 35, 02206

\bibitem[{McKay {et~al.}(1991)McKay, Pollack, \& Courtin}]{mckay1991greenhouse}
McKay, C., Pollack, J., \& Courtin, R. 1991, Science, 253, 1118

\bibitem[{Mitchell(2008)}]{mitchell2008drying}
Mitchell, J. 2008, Journal of Geophysical Research, 113, E08015

\bibitem[{Mitchell {et~al.}(2011)Mitchell, {\'A}d{\'a}mkovics, Caballero, \&
  Turtle}]{mitchell2011locally}
Mitchell, J., {\'A}d{\'a}mkovics, M., Caballero, R., \& Turtle, E. 2011, Nature
  Geoscience, 4, 589

\bibitem[{Mitchell {et~al.}(2006)Mitchell, Pierrehumbert, Frierson, \&
  Caballero}]{mitchell2006dynamics}
Mitchell, J., Pierrehumbert, R., Frierson, D., \& Caballero, R. 2006,
  Proceedings of the National Academy of Sciences, 103, 18421

\bibitem[{Mitchell {et~al.}(2009)Mitchell, Pierrehumbert, Frierson, \&
  Caballero}]{mitchell2009impact}
---. 2009, Icarus, 203, 250

\bibitem[{Rannou {et~al.}(2006)Rannou, Montmessin, Hourdin, \&
  Lebonnois}]{rannou2006latitudinal}
Rannou, P., Montmessin, F., Hourdin, F., \& Lebonnois, S. 2006, Science, 311,
  201

\bibitem[{Rodriguez {et~al.}(2011)Rodriguez, Le~Mou{\'e}lic, Rannou, Sotin,
  Brown, Barnes, Griffith, Burgalat, Baines, Buratti,
  {et~al.}}]{rodriguez2011titan}
Rodriguez, S., Le~Mou{\'e}lic, S., Rannou, P., Sotin, C., Brown, R., Barnes,
  J., Griffith, C., Burgalat, J., Baines, K., Buratti, B., {et~al.} 2011,
  Icarus, 216, 89

\bibitem[{Schaller {et~al.}(2006)Schaller, Brown, Roe, \&
  Bouchez}]{Schaller_etal06}
Schaller, E., Brown, M., Roe, H., \& Bouchez, A. 2006, Icarus, 182, 224

\bibitem[{Schneider {et~al.}(2012)Schneider, Graves, Schaller, \&
  Brown}]{schneider2012polar}
Schneider, T., Graves, S., Schaller, E., \& Brown, M. 2012, Nature, 481, 58

\bibitem[{Trenberth \&
  Stepaniak(2003{\natexlab{a}})}]{trenberth2003covariability}
Trenberth, K., \& Stepaniak, D. 2003{\natexlab{a}}, Journal of climate, 16,
  3691

\bibitem[{Trenberth \& Stepaniak(2003{\natexlab{b}})}]{trenberth2003seamless}
---. 2003{\natexlab{b}}, Journal of Climate, 16, 3706

\bibitem[{Turtle {et~al.}(2011{\natexlab{a}})Turtle, Del~Genio, Barbara, Perry,
  Schaller, McEwen, West, \& Ray}]{turtle2011seasonal}
Turtle, E., Del~Genio, A., Barbara, J., Perry, J., Schaller, E., McEwen, A.,
  West, R., \& Ray, T. 2011{\natexlab{a}}, Geophysical Research Letters, 38,
  L03203

\bibitem[{Turtle {et~al.}(2011{\natexlab{b}})Turtle, Perry, Hayes, Lorenz,
  Barnes, McEwen, West, Del~Genio, Barbara, Lunine, {et~al.}}]{Turtle_etal11}
Turtle, E., Perry, J., Hayes, A., Lorenz, R., Barnes, J., McEwen, A., West, R.,
  Del~Genio, A., Barbara, J., Lunine, J., {et~al.} 2011{\natexlab{b}}, science,
  331, 1414

\bibitem[{Turtle {et~al.}(2009)Turtle, Perry, McEwen, DelGenio, Barbara, West,
  Dawson, \& Porco}]{Turtle_etal09}
Turtle, E., Perry, J., McEwen, A., DelGenio, A., Barbara, J., West, R., Dawson,
  D., \& Porco, C. 2009, Geophys. Res. Lett, 36, L02204

\bibitem[{{Williams} {et~al.}(2012){Williams}, {McKay}, \&
  {Persson}}]{williams2012energy}
{Williams}, K.~E., {McKay}, C.~P., \& {Persson}, F. 2012, Planetary and Space
  Science, 60, 376

\end{thebibliography}
\end{document}